\documentclass[preprint, double-spaced,showpacs,preprintnumbers,amsmath,amssymb,prb]{revtex4}
\usepackage[ansinew]{inputenc}
\usepackage{color}
\usepackage{longtable}
\usepackage{xspace,graphicx}

\newcommand{\NH}{\hbox{NH$_3$}\xspace}
\newcommand{\CH}{\hbox{CH$_4$}\xspace}
\newcommand{\ND}{\hbox{ND$_3$}\xspace}
\newcommand{\hho}{\hbox{H$_2$O}\xspace}
\newcommand{\PDEUN}{\hbox{P2$_{1}$2$_{1}$2$_{1}$}\xspace}
\newcommand{\etal}{\emph{et al.}\xspace}

\graphicspath{{./fig/}}

\begin{document}

\title{High pressure-high temperature phase diagram of ammonia}
\author{S. Ninet}
\email{sandra.ninet@impmc.jussieu.fr}
\author{F. Datchi}
\affiliation{Institut de Minéralogie et de Physique des Milieux Condensés, Département Physique des Milieux
Denses, CNRS UMR 7590, Université Pierre et Marie Curie - Paris VI, 4 place Jussieu, 75252 Paris Cedex 05,
France.}

\date{\today}

\begin{abstract}
The high pressure(P)-high temperature(T) phase diagram of solid ammonia has been investigated using diamond anvil
cell and resistive heating techniques. The III-IV transition line has been determined up to 20 GPa and 500 K both
on compression and decompression paths. No discontinuity is observed at the expected location for the III-IV-V
triple point. The melting line has been determined by visual observations of the fluid-solid equilibrium up to 9
GPa and 900 K. The experimental data is well fitted by a Simon-Glatzel equation in the covered P-T range. These
transition lines and their extrapolations are compared with reported \emph{ab initio} calculations.
\end{abstract}


\maketitle


\section{Introduction}

Giant planets are mainly composed of simple molecular compounds such as H$_{2}$O, H$_{2}$, He, CH$_{4}$ and
\NH\cite{Hubbard1981}. In particular, a layer of mixed ices (\NH, H$_{2}$O, CH$_{4}$) has been proposed to exist
in the interior of Uranus and Neptune. In this layer, extreme thermodynamic conditions are supposed (20$<$P$<$300
GPa and 2000$<$T$<$5000~K)\cite{Hubbard1995} and the knowledge of the properties of these "hot" ices are crucial
to a good examination of astrophysical data.

Experimental data under high P-T conditions are also very important to validate \emph{ab initio}
calculations\cite{Cavazzoni1999,Goldman2005} performed on \NH and H$_{2}$O. A very similar phase diagram has been
predicted for these two compounds. In particular, a spectacular superionic phase is predicted at high pressures
and temperatures\cite{Cavazzoni1999}. In this phase, the molecular nature of these ices disappears and hydrogen
atoms diffuse in all the N- or O- network. This prediction has stimulated many recent experimental works on the
melting curve of water as the latter is expected to exhibit a clear kink at the onset of the transition. A change
of slope in the melting curve of water ice has been reported in three different experimental studies, but its
location substantially differ between them: 47 GPa and 1000~K in Ref.~\onlinecite{Goncharov2005}, 43 GPa and
1500~K in Ref. \onlinecite{Schwager2004}, and 35 GPa and 1040~K in Ref.~\onlinecite{Lin2005}. By contrast, no
discontinuity has been observed in another melting study up to 50 GPa and 1100~K~\cite{Dubrovinskaya2004}.
Theoretical predictions of the fluid-superionic solid transition pressure range between 30 and 75 GPa at
2000~K~\cite{Cavazzoni1999,Goldman2005}. The exact location of the superionic phase is important for models of
planetary interiors as it determines for example whether the Neptune and Uranus isentropes cross the superionic
phase.

Unlike water, experimental data on ammonia under high static P-T conditions are very scarce. Experiments in a
diamond anvil cell have been so far limited to 373~K~\cite{Hanson1980}. Several shock-wave
experiments~\cite{Dick1981,Mitchell1982,Nellis1988,Radousky1990,Nellis1997} have reported equation of state and
electrical conductivity data in the range $\sim$2-65 GPa, 1100-4600~K, but these are restricted to the fluid
phase. New experiments are thus needed to bridge the present gap between static and dynamic investigations.

\begin{figure}
\includegraphics[width=3in]{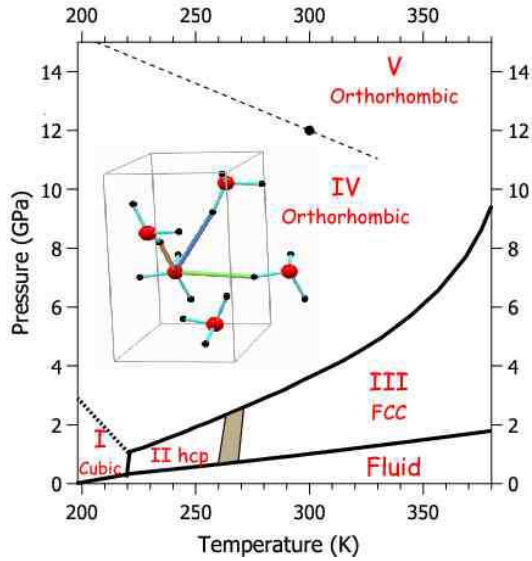}
\caption{\label{PhaseDiagram} (Color online) Phase diagram of ammonia (\NH). The structure of phase IV as
determined by Loveday \emph{et al}\cite{Loveday1996} is depicted. }
\end{figure}

The presently accepted phase diagram of ammonia is shown in Fig.~\ref{PhaseDiagram}. Above 4 GPa, three different
phases may be stabilized depending on the temperature. In increasing order of temperature, these are the ordered,
orthorhombic phase IV, the plastic (orientationally disordered) cubic phase III, and the fluid. The IV-III and
III-fluid transition lines have been determined up to 373~K~\cite{Hanson1980,Mills1982,Kume2001}. Evidences for a
solid phase transition  in \NH at 14 GPa and 300~K to phase V were initially reported by Gauthier \etal using
Raman~\cite{Gauthier1988} and Brillouin~\cite{Gauthier1988b} spectroscopies. We recently confirmed this
transition using single-crystal x-ray diffraction~\cite{Datchi2006} and Raman spectroscopy\cite{NinetTBP}, but
unlike Gauthier \etal's suggestion of a cubic structure for the high pressure phase, we found that the latter is
isostructural to phase IV (space group \PDEUN). The transition was detected at 12 and 18 GPa for \NH and \ND
respectively at room temperature.

In this paper, we report experimental measurements of the phase diagram of \NH in the range [1-20 GPa] and
[300-900K]. The III-IV transition line has been determined up to 20 GPa and 500 K in order to detect the
influence of the isostructural phase transition. The melting curve has been determined up to 9~GPa and 905~K. No
evidence for the superionic phase has been observed in this P-T range.

\section{Experimental methods}

Samples of \NH were cryogenically loaded in membrane diamond anvil cells as described in
Refs.~[\onlinecite{Datchi2006,Ninet2006}]. A gold ring between the rhenium gasket and the sample was employed to
prevent any possible chemical reaction between ammonia and rhenium at high temperature. Pressure was determined
with the  luminescence of SrB$_{4}$O$_{7}$:Sm$^{2+}$ at high temperature\cite{Datchi1997,Datchi2007}. A ruby ball
was also used below 600 K. The uncertainties on pressure measurements are typically less than 0.02 GPa at ambient
temperature and around 0.15 GPa at 900~K.

Membrane diamond anvil cells (mDAC) made of high-temperature Inconel alloy were used. With a commercial
ring-shaped resistive heater around the mDAC, sample temperatures up to 750~K could be obtained. Above 750~K, an
internal heater located around the diamond-gasket assembly is used in addition. These heaters are temperature
regulated within 1~K. During heating, a continuous flow of Ar/H$_{2}$ reducing gas mixture is directed onto the
mDAC. The temperature of the sample was determined thanks to a K-type thermocouple fixed by ceramic cement on the
head of one diamond, and cross-checked with the temperature determined \emph{in situ} from the ruby up to
600~K~\cite{Datchi2007}. With these techniques, a temperature uncertainty inferior to 5~K is obtained, as
validated by previous studies~\cite{Datchi2000,Datchi2004,Giordano2007}.

Measurements of Raman spectra were performed using a T64000 spectrometer in backscattering geometry.
Angular-dispersive x-ray diffraction spectra were collected on the ID09 station of the ESRF (Grenoble, France).

\begin{figure}
\includegraphics[width=3in]{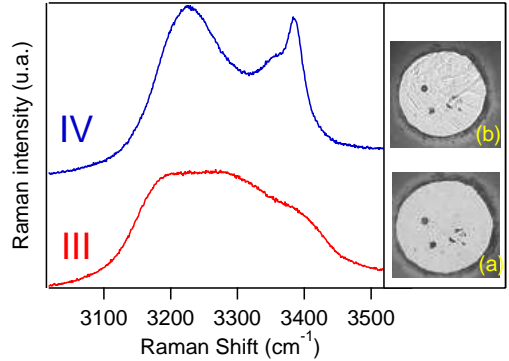}
\caption{\label{RamanSpectra}Raman spectra of solid \NH phases IV and III in the $\nu_1$-$\nu_3$ region. The
spectra were collected at 373~K, 9.7 and 9 GPa.  Photographs of polycrystalline samples of phase III (a) and IV
(b) are shown in inset. }
\end{figure}

\section{Results and discussion}

\subsection{III-IV transition line}
The transition between the disordered phase III and ordered phase IV has been investigated as a function of
pressure and temperature up to 20 GPa and 500~K. The transition pressure has been determined along isotherms
during compression and decompression. We used several criteria to detect the transition: (1) by detecting the
pressure discontinuity: a rapid pressure drop of a few tenth of GPa is observed at the III-IV transition due to
the volume discontinuity (1.5\% at 300~K\cite{Datchi2006}). The transition pressure is defined as the lower
pressure; (2) by X-ray diffraction: patterns of phase III (cubic) and phase IV (orthorhombic) are easily
recognizable; (3) by visual observation: reticulation is observed in phase IV (birefringent phase), which
disappears in  phase III (see pictures in Fig.~\ref{RamanSpectra}); (4) by Raman spectroscopy: the shape of the
$\nu_1-\nu_3$ band is very different between the two phases; the modes are broad in the disordered phase III and
sharper in the ordered phase IV (Fig.~\ref{RamanSpectra}).

\begin{figure}
\includegraphics[width=3.4in]{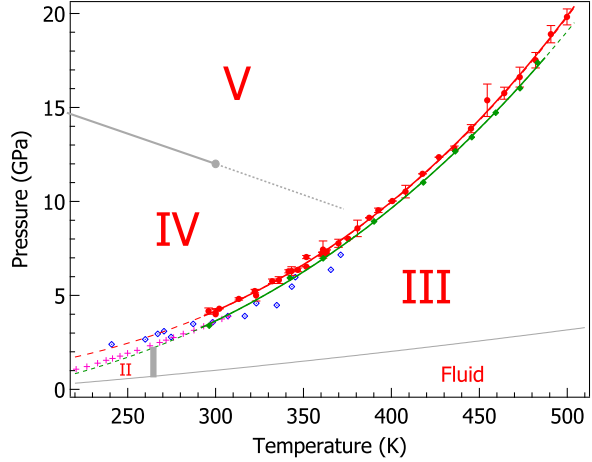}
\caption{\label{lineIII-IV}The III-IV transition line. Our experimental points in compression (dots) and
decompression (green lozenge) are presented along with literature data (crosses: Kume~\etal\cite{Kume2001},
diamonds: Hanson~\etal\cite{Hanson1980}). Solid lines represent the fits to Simon-Glatzel equations
(Eqs.~[\ref{simoncomp},\ref{simondecomp},\ref{simonmoyen}])}
\end{figure}

Our experimental points are presented in Fig.~\ref{lineIII-IV} and in Table~\ref{data_III_IV}. They form two
distinct transition curves corresponding respectively to compression and decompression measurements. These two
data sets are well fitted by the following Simon-Glatzel (SG) type equations\cite{Simon1929} :

\begin{equation}\label{simoncomp}
    P_{III\rightarrow
IV}(T)=3.9(1)+3.7(2)\left[\left(\frac{T}{293.15}\right)^{3.1(1)}-1\right]
\end{equation}

\begin{equation}\label{simondecomp}
    P_{IV\rightarrow
III}(T)=3.3(1)+4.5(4)\left[\left(\frac{T}{293.15}\right)^{2.8(1)}-1\right]
\end{equation}

Taking the average of these two curves as the transition line gives the following SG equation:
\begin{equation}\label{simonmoyen}
    P_{IV\rightleftarrows
III}(T)=3.6(2)+4.1(6)\left[\left(\frac{T}{293.15}\right)^{3.0(2)}-1\right]
\end{equation}

 The use of the SG equation is usually restricted to the description of melting curves. We note though that this
phenomenological law is valid for any first-order phase transition satisfying Clausius-Clapeyron's relation
($dP/dT=\Delta H/T\Delta V$ where $\Delta H$ and $\Delta V$ are the enthalpy and volume discontinuities at the
transition), provided that $\Delta H/\Delta V$ is a linear function of pressure~\cite{Voronel1958}. The III-IV
transition is a first-order transition between a plastic phase (III) where the H atoms are highly disordered
(they may occupy as much as 192 positions \cite{VonDreele1984}) and an ordered phase (IV) where H atoms have
definite positions. The fcc packing in phase III also indicates the weak influence of the H-bonds in this phase.
This phase transition can thus be viewed as a pseudo-melting of the hydrogen sublattice.

\begin{table}
\caption{\label{data_III_IV} Experimental data for the III-IV transition line. T is the temperature measured with
the thermocouple in K. P is the pressure measured with the borate or the ruby in GPa.}
\begin{ruledtabular}
\begin{tabular}{cc|cc|cc}
   P\emph{$_{III\rightarrow IV}$} & T$_{III\rightarrow IV}$   & P$_{III\rightarrow IV}$ & T$_{III\rightarrow IV}$    &P$_{III\rightarrow IV}$ & T$_{III\rightarrow IV}$    \\
  \hline
  3.69 & 291.0   & 7.33 & 363.0    &  13.90 & 445.3  \\
  4.00 & 298.0   & 7.47 & 361.1  &  14.92 & 454.5  \\
  4.08 & 298.0   & 7.66 & 369.9  &  15.41 & 454.5  \\
  4.16 & 298.0   & 7.80 & 369.9  &  15.78 & 464.2  \\
  4.29 & 302.1   & 8.00 & 375.5  &  16.64 & 473.0    \\
  4.52 & 313.5   & 8.04 & 375.5  &  16.71 & 473.0    \\
  4.81 & 313.2   & 8.59 & 380.6  &  17.54 & 482.0    \\
  5.03 & 323.0   & 9.05 & 384.4  &  18.92 & 490.7  \\
  5.06 & 323.0   & 9.16 & 387.4  &  19.84 & 499.9  \\
  5.23 & 322.3   & 7.33 & 363.0   &        &        \\
  5.75 & 332.2   & 7.47 & 361.1  &        &        \\
  5.79 & 336.0   & 7.66 & 369.9  &        &        \\
  5.84 & 336.0   & 7.80 & 369.9  &         &       \\
  5.93 & 328.8   & 8.00 & 375.5  &        &        \\
  6.24 & 343.2   & 8.04 & 375.5  &        &        \\
  \cline{5-6}
  6.28 & 341.5   & 8.59 & 380.6  & P$_{IV\rightarrow III}$ & T$_{IV\rightarrow III}$   \\
  \cline{5-6}
  6.32 & 343.2   & 9.05 & 384.4  & 17.37 & 483.1   \\
  6.55 & 351.6   & 9.16 & 387.4  & 16.02 & 473.1   \\
  6.58 & 351.6   & 9.55  & 392.5 & 14.72 & 459.3   \\
  6.36 & 346.6   & 9.89  & 400.5 & 13.42 & 445.7   \\
  7.04 & 351.6   & 10.05 & 400.5 & 12.66 & 436.3   \\
  7.18 & 360.5   & 10.39 & 408.0 & 11.00 & 418.1   \\
  7.22 & 360.9   & 10.55 & 408.0 & 8.95  & 390.1   \\
  7.24 & 361.1   & 11.50 & 417.7 & 6.97  & 361.1   \\
  7.26 & 361.9   & 12.21 & 427.0 & 5.93  & 342.2   \\
  7.31 & 362.7   & 12.68 & 435.8 & 3.39  & 296.3   \\
  7.31 & 361.1   & 12.86 & 435.8 &       &         \\
\end{tabular}
\end{ruledtabular}
\end{table}

The experimental data points of Hanson and Jordan~\cite{Hanson1980} and Kume \etal~\cite{Kume2001} are also
presented in figure~\ref{lineIII-IV}. Hanson and Jordan's transition pressures above 300~K are on average lower
than ours but the slopes are very similar. A good agreement can also be observed between Kume \etal's data and
our experimental points taken on decompression.

Our main goal in studying the III-IV transition line was to detect whether the presence of the transition between
the two isostructural solids IV and V detected at 12 GPa at 300~K  had a measurable inference on this line, such
as a discontinuity of slope. This is actually not the case, no discontinuity being observable within the
precision of our measurements up to 500~K and 20 GPa. In our single-crystal x-ray diffraction experiments, the
IV-V transition was detected thanks to the systematic and sudden splitting of the crystal and the discontinuous
change of slope of the $c/a$ ratio~\cite{Datchi2006}. Although no volume jump could be measured within
uncertainties ($\sim0.06$ cm$^3$/mol), the transition must be first-order since the two phases are isosymmetric.
Using Raman spectroscopy, we observed a small jump in some lattice modes at 12 GPa, 300~K and around 20 GPa at
50~K \cite{NinetTBP}. In figure~\ref{lineIII-IV}, the IV-V transition line based on these two points is drawn.
The absence of discontinuity at the expected III-IV-V triple point (ca. 9 GPa, 390 K) is not really surprising
since the volume difference between IV and V is  very small. It is also possible that, between 300 and 380~K, the
IV-V transition line ends at a critical point where the transition becomes second-order. The existence of such
critical points have been predicted by Landau \cite{Toledano1987,Landau} and observed in a few materials such as
Cr-doped V$_2$O$_3$\cite{Jayaraman1970}, NH$_{4}$PF$_{6}$\cite{Swainson2002} and cerium\cite{Beecroft1960}. The
proximity of a Landau critical point could also explain the weak volume discontinuity at ambient temperature.

\subsection{The melting line}

\begin{figure}
\includegraphics[width=3.7in]{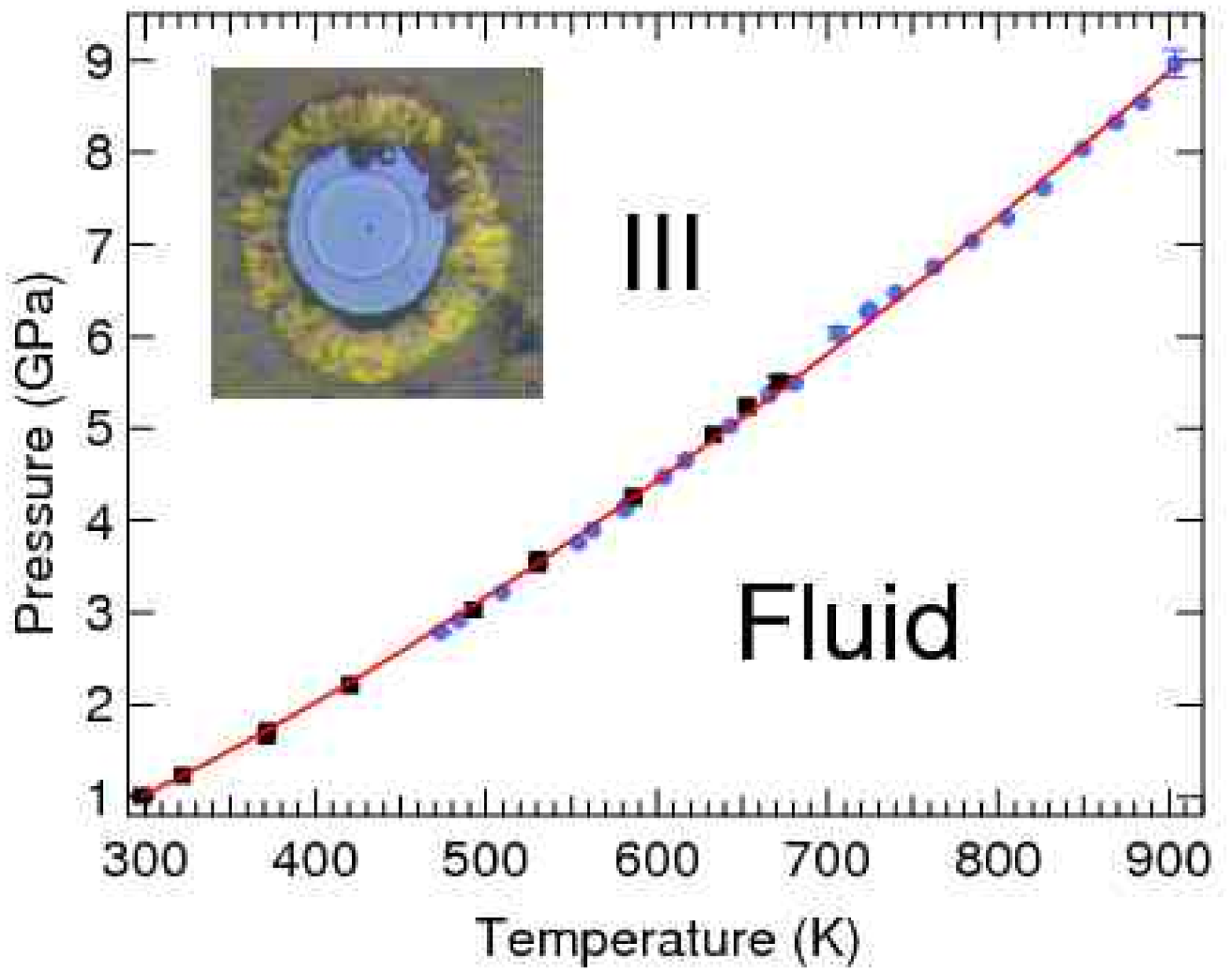}
\caption{\label{melting}Experimental melting points determined in this work. The symbols are associated with two
separates experiments. The solid line represents the fit to our data with a Simon-Glatzel equation. The inset
shows a photograph of the solid-fluid equilibrium at 473~K. A ruby ball and some samarium powder are visible. }
\includegraphics[width=3.2in]{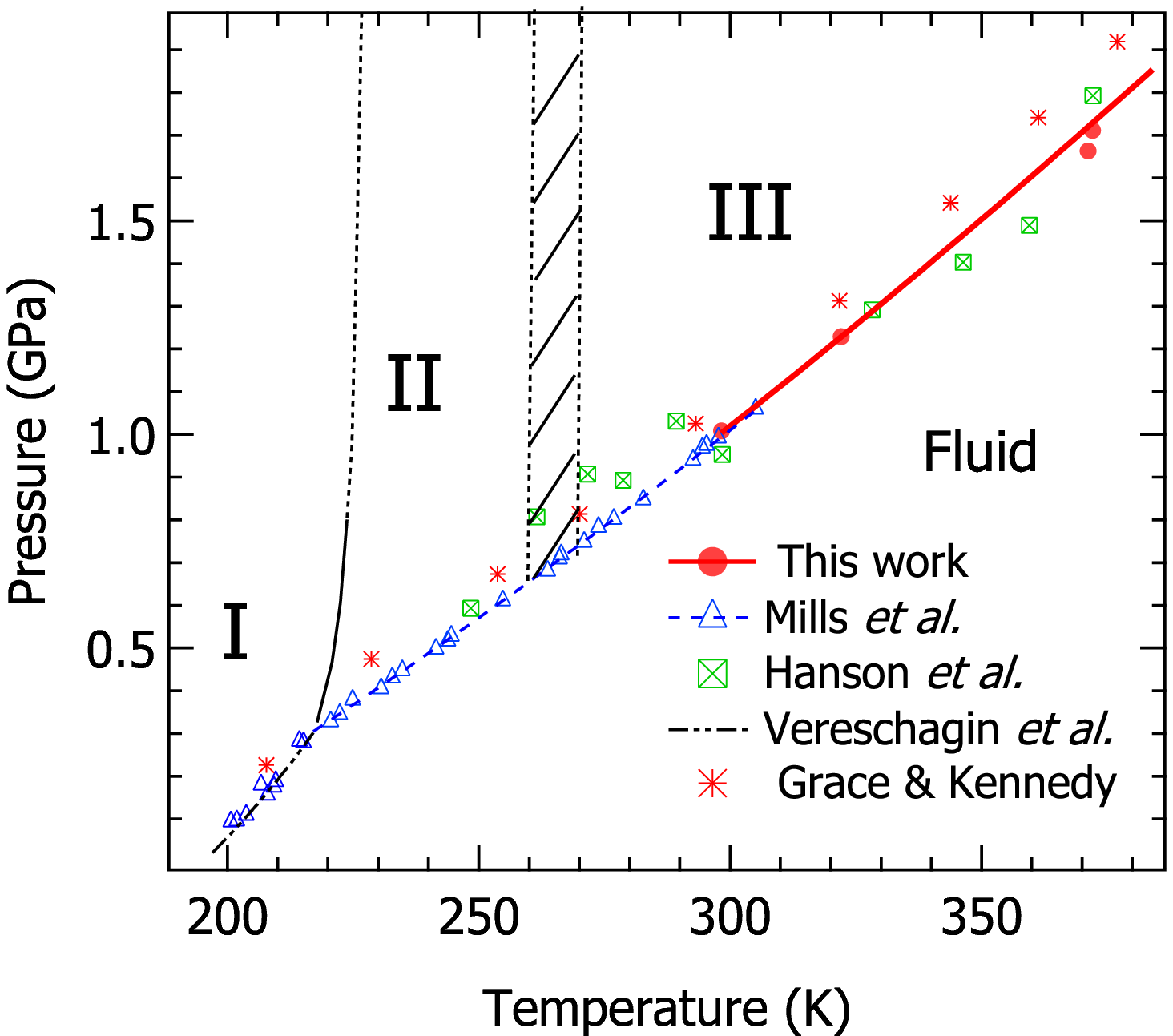}
\caption{\label{melting_lit} Comparison of the melting line obtained in this work (red dots and line) with
previous experimental studies\cite{Hanson1980,Mills1982,Vereschagin1958,Grace1967}.}
\end{figure}

The melting curve was determined by visual observation of the solid-fluid equilibrium. This is possible because
of the difference in refractive index between the fluid and solid phases which remains large enough to clearly
distinguish the two phases up to 900~K (see inset of Fig.~\ref{melting}). The measurement of the sample pressure
and temperature when this coexistence has been stabilized defines a melting point. The coexistence was kept
during the slow heating of the sample by increasing the load. This method allows to precisely determine  the
melting curve and prevents problems associated with metastabilities such as undercooling or overpressurization.

\begin{table}
\caption{\label{data_melt} Experimental data for the melting line. $T_m$ is the temperature measured with the
thermocouple in K. $P_m$ is the pressure measured with the borate in GPa.}
\begin{ruledtabular}
\begin{tabular}{cc@{\hspace{1.2cm}}|cc}
   $P_{m}$ & $T_{m}$ & $P_{m}$ & $T_{m}$\\
  \hline
  1.01 & 298.3  & 4.94 & 6 33.0  \\
  1.23 & 322.0  & 5.03 & 6 43.1  \\
  1.66 & 371.1  & 5.24 & 6 52.9  \\
  1.71 & 372.0  & 5.37 & 6 65.8  \\
  2.21 & 420.1  & 5.50 & 681.1  \\
  2.79 & 473.1  & 5.51 & 671.7  \\
  2.93 & 484.3  & 6.04 & 705.9  \\
  3.03 & 492.8  & 6.26 & 723.4  \\
  3.22 & 509.4  & 6.30 & 724.6  \\
  3.53 & 529.9  & 6.47 & 740.1  \\
  3.57 & 530.4  & 6.76 & 762.6  \\
  3.76 & 553.8  & 7.04 & 784.5  \\
  3.90 & 562.1  & 7.29 & 805.1  \\
  4.12 & 580.8  & 7.62  & 826.4  \\
  4.15 & 582.0  & 8.04 & 849.6  \\
  4.26 & 586.3  & 8.33  & 869.0  \\
  4.48 & 604.1  & 8.55 & 884.3  \\
  4.65 & 616.8  & 8.97 & 903.9  \\
\end{tabular}
\end{ruledtabular}

\end{table}

Two different samples have been studied to determine the melting curve of \NH in the range [300-904~K] and [1-9
GPa]. The two measurements agree very well in the overlapping region (470-670~K). The experimental melting points
are plotted in Fig.~\ref{melting} (reported in table~\ref{data_melt}) and compared to other melting
studies\cite{Hanson1980, Mills1982,Vereschagin1958,Grace1967} in Fig.~\ref{melting_lit}. Our data agree with
previous measurements from Hanson and Jordan \cite{Hanson1980} up to 373~K within their stated uncertainty of 0.1
GPa. The melting pressures determined by Grace and Kennedy~\cite{Grace1967} in a piston-cylinder apparatus are
systematically higher than ours and those of Mills \etal~\cite{Mills1982} by 0.08--0.13 GPa. In the P-T range
covered by our experiments, the melting pressure is a monotonous increasing function of temperature and no
discontinuity is observed. The whole data set is well fitted by the following Simon-Glatzel equation
\footnote{Numbers in parentheses indicate the 95\% confidence interval for the fit parameters}:

\begin{equation}\label{simon}
    P(T)=0.307+1.135(51)\left[\left(\frac{T}{217.34}\right)^{1.510(31)}-1\right]
\end{equation}

In the latter expression, we used as reference P-T point the I-II-fluid triple point coordinates of \NH
(P=0.307~GPa and T=217.34~K)~\cite{Mills1982}. As a matter of fact, the II-III-fluid triple point has not been
determined yet: no discontinuity has been observed on the melting curve~\cite{Mills1982}  at the expected
location for this triple point (around 265~K\cite{Millsunpub}). Phase II differs from phase III by its hexagonal
compact ordering of the molecules, but both are plastic phases with large orientational disorder. The absence of
discontinuity observable on the melting curve at the triple point indicates that the two phases have  very
similar free energies. The extrapolation of our melting curve down to the I-II-triple point reproduces very well
the melting data of Mills \etal~\cite{Mills1982}. Actually, the parameters of the Simon-Glatzel form in
Eq.~\ref{simon}, obtained by fitting our data alone, are identical within standard deviations to those determined
by Mills \etal~\cite{Mills1982} by fitting their own melting data between 220.5 and 305 K.

\begin{figure}
\includegraphics[width=2.8in]{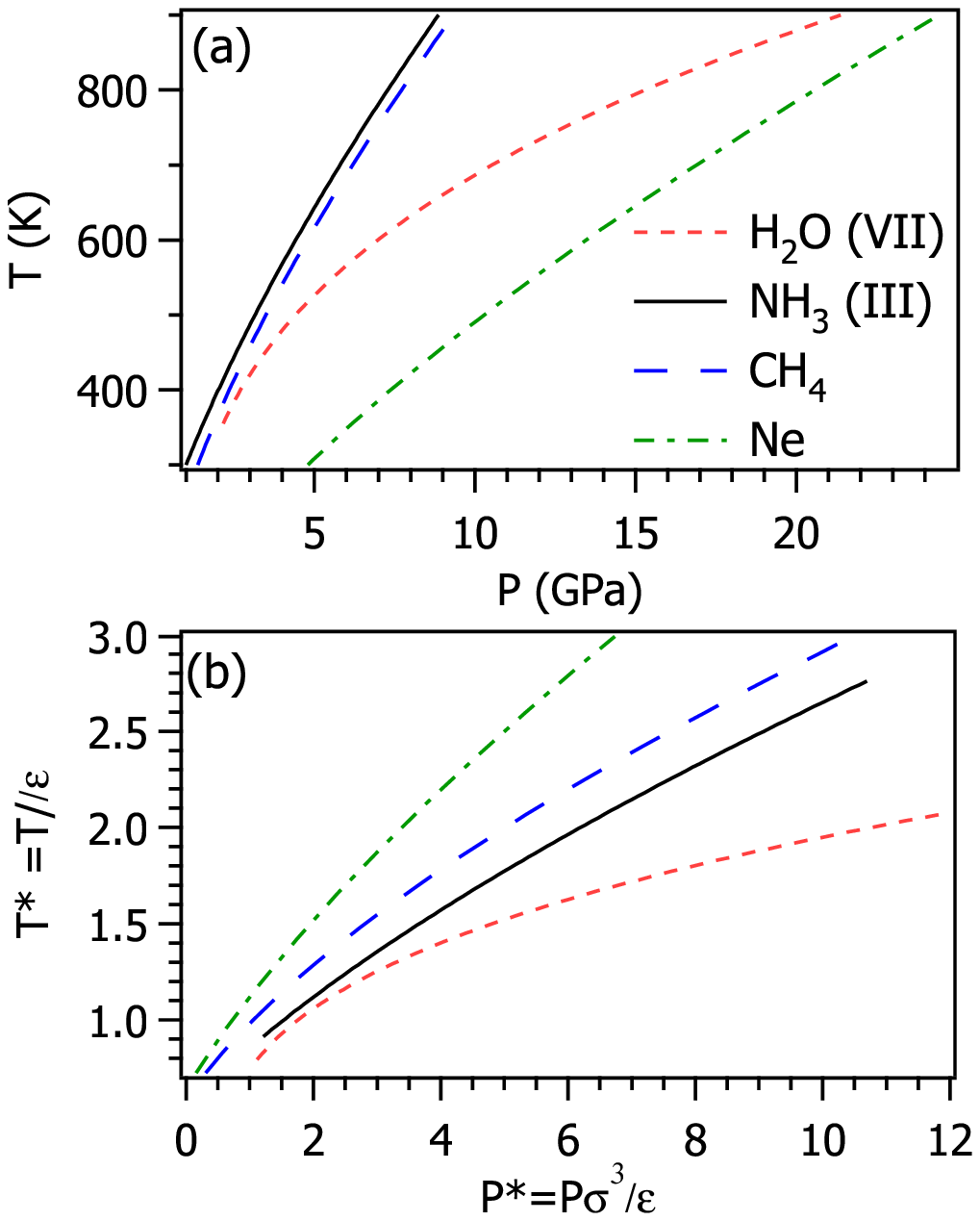}
\caption{\label{isocomp}(Color online) Comparison between the melting lines of the isoelectronic solids \hho,
\NH, CH$_4$ and Ne on absolute (a) and reduced (b) scales.}
\end{figure}


The melting line of \NH is compared to those of the isoelectronic solids \hho~\cite{Datchi2000},
CH$_4$\footnote{The melting line of methane has only been determined up to 400~K. At higher temperatures, we use
the extrapolation of the melting law given in Ref.~\onlinecite{Setzmann1991}} and neon\cite{Vos1991, Datchi2004}
in Fig.~\ref{isocomp}. This comparison is shown both on a absolute P-T scale and using reduced units\footnote{We
use reduced units derived from the law of corresponding states : $P^*=P\sigma^3/\epsilon$, $T^*=T/\epsilon$,
where $\sigma$ and $\epsilon$ are respectively the position and depth of the mimimum of the effective pair
potential. For these two parameters, we use the values determined for the exp-6 potential by Belonoshko and
Saxena~\cite{Belonoshko1992} for \hho, \NH and CH$_4$, and by Vos \etal~\cite{Vos1991} for Ne.} to rescale the
melting lines on the same density map.  Although none of these solids "corresponds" \emph{stricto sensu},  it can
be seen that the melting line of \NH is closer to that of methane than to the one of water at high P-T. In
particular, the slopes $dT_m/dP_m$ of the melting curves of \NH and \CH are very similar and much less pressure
dependent than for \hho. Since the main difference between \CH and \hho, in terms of intermolecular interactions,
is the absence of hydrogen bonds in \CH, the similitude between \CH and \NH indicates that the hydrogen bonds
have little influence on the melting properties of \NH in the P-T range covered by our experiments.

\subsection{Phase diagram at high temperatures: comparison with \emph{ab initio} calculations}

\begin{figure}
\includegraphics[width=3.7in]{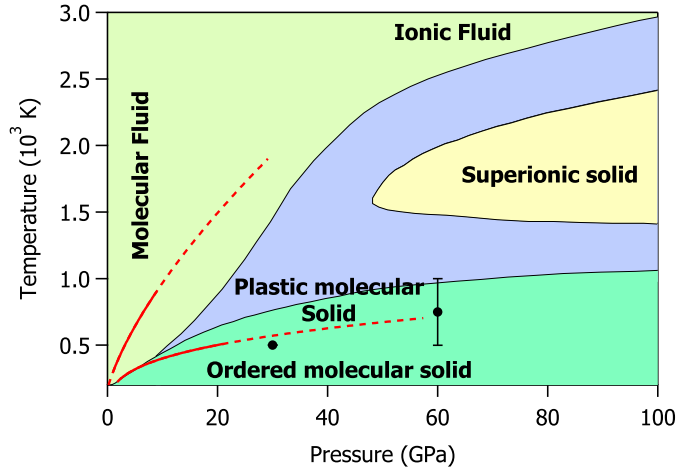}
\caption{\label{comp_calc}(Color online) Phase diagram of ammonia according to \emph{ab initio} molecular
dynamics simulations of Ref.~\onlinecite{Cavazzoni1999}. Fits to present experimental data (red solid lines) and
their extrapolations (broken lines) are shown. The black dots represent P-T points where simulations predict a
transition between phase IV and a hcp plastic solid.}
\end{figure}

To compare our experimental results to the \emph{ab initio} molecular dynamics (AIMD) simulation of
Cavazzoni~\etal\cite{Cavazzoni1999}, we reproduced the phase diagram predicted by these authors in
Fig.~\ref{comp_calc}. Since the calculations probed pressures above 30 GPa, comparison can only be made with the
extrapolation of the Simon-Glatzel equations determined for both the III-IV phase line (Eq. [\ref{simonmoyen}])
and the melting line (Eq. [\ref{simon}]). Starting from a sample of phase IV at 30 GPa, Cavazzoni~\etal observed
a phase transition to a hcp plastic phase at ca. 500~K and then melting around 1500~K.  At 60 GPa, the
ordered-plastic phase transition was observed between 500 and 1000~K, succeeded by the plastic-superionic phase
transition between 1000~K and 1200~K. This superionic phase was also obtained at 150 and 300 GPa in the same
temperature range. Although phase III is fcc, we have observed above that the difference in free energy between
the hcp and fcc plastic phases is very small, so we can assimilate the hcp plastic phase obtained by Cavazzoni
\etal with phase III. The extrapolation of our III-IV transition line puts the transition at 570~K at 30 GPa and
720~K at 60 GPa, i.e. close to the predicted ones. On the other hand, the experimental and calculated melting
lines are rather different. Cavazzoni~\etal predict a melting temperature of $\sim$1500~K at 30 GPa, that is
400~K lower than the extrapolation of our Simon-Glatzel fit. It is rather surprising since AIMD simulations
usually tend to overestimate the melting temperature~\cite{Koci2007}. If we rely on calculations, a strong
deviation from the behaviour predicted by the Simon-Glatzel equation should be observed between 9 and 30~GPa.
This is one motivation to pursue the experiments to higher P-T conditions.

\section{Conclusions}

In this paper, we have presented an experimental investigation of the high P-T phase diagram of \NH. We have
examined the evolution of the III-IV transition line up to 500~K and of the melting curve up to 900 K. These two
first-order transition lines can be well fitted with Simon-Glatzel equations and no discontinuities have been
observed on both lines. The presence of a Landau critical point ending the IV-V first-order transition line is a
possible explanation for the non-observation of the III-IV-V triple point. A good agreement is found between the
extrapolation of present measurements and \emph{ab initio} predictions of the ordered-plastic solid transitions,
but not for the melting curve. Higher P-T conditions need to be probed to evidence the predicted superionic
phase.

\begin{acknowledgments}
The authors are indebted to B. Canny for his help in preparing the experiments and acknowledge the ESRF for
provision of beamtime.
\end{acknowledgments}

\end{document}